\documentclass[showpacs,preprintnumbers,prd,nofootinbib,floats,amssymb,floatfix]{revtex4}
\usepackage{amsmath}
\usepackage{amsfonts}
\usepackage{graphicx}
\usepackage{hyperref}
\usepackage{amsmath}
\usepackage{amstext}
 
\begin{document}

\title { Hamiltonian Dynamics for Proca's theories    in five dimensions with a compact dimension }
%\end{flushleft}
 \author{ Alberto Escalante}  \email{aescalan@ifuap.buap.mx}
 \author{Carlos L. Pando Lambruschini} 
% \author{Mois\'es Z\'arate}
%\author{ Moises Zarate}  \email{mzarate@ifuap.buap.mx}
\affiliation{ Instituto de F{\'i}sica Luis Rivera Terrazas, Benem\'erita Universidad Aut\'onoma de Puebla, (IFUAP). Apartado
 postal J-48 72570 Puebla. Pue., M\'exico,}
 \author{ Prihel Cavildo }  %\email{}
\affiliation{ Instituto de F{\'i}sica Luis Rivera Terrazas, Benem\'erita Universidad Aut\'onoma de Puebla, (IFUAP). Apartado postal J-48 72570 Puebla. Pue., M\'exico,}
 \affiliation{ Facultad de Ciencias F\'{\i}sico Matem\'{a}ticas, Benem\'erita Universidad Au\-t\'o\-no\-ma de Puebla,
 Apartado postal 1152, 72001 Puebla, Pue., M\'exico.}

\begin{abstract} 
The canonical analysis of  Proca's  theory in five dimensions with a compact dimension is performed. From the Proca five dimensional action, we perform  the compactification process   on a $S^1/\mathbf{Z_2}$ orbifold,  then,  we analyze the   four dimensional effective action that emerges from the compactification process. We report   the extended action, the extended Hamiltonian and the counting of degrees of freedom of the theory. We show that the theory with  the compact dimension   continues laking of first class constraints. In fact, the final theory is not a gauge theory and  describes the propagation of   a  massive vector field  plus a tower of  massive $KK$-excitations and one massive  scalar field. Finally, we perform the analysis of a  5D $BF$-like theory plus a Proca's term, we  perform the compactification process and we find all  constraints of the theory,  we also carry out  the counting of physical degrees of freedom; with these results, we conclude that the theory is not topological  and has    reducibility conditions among the constraints. 
\end{abstract}

\date{\today}
\preprint{}
\maketitle
\section{INTRODUCTION}
\vspace{1em} \
Nowadays, the introduction of extra dimensions  in field theories have allowed a  new way of
 looking at  several problems in  theoretical physics. It is well-know that the first  proposal  introducing extra dimensions beyond the fourth dimension was considered around 1920's, when   Kaluza and Klein (KK) tried to unify electromagnetism 
 with Einstein's gravity by proposing a theory in 5D  where the fifth dimension is compactified on a  circle  $S^1$ of radius $R$, and 
 the electromagnetic field  is   contained as a  component of the  metric tensor  \cite{1}. The study of models involving extra dimensions  has an   important activity  in order to explain and solve 
 some  fundamental  issues  found in theoretical physics,  such as,  the problem of  mass hierarchy,   the explanation of dark energy,  dark matter and inflation  etc., \cite{1a}. Moreover, extra dimensions become also  important  in theories of  grand unification trying  of incorporating   gravity and gauge interactions in a theory of everyting. In this respect, it is well known that extra dimensions have a  fundamental role in the developing of  string theory, since all versions of the theory  are formulated  in a spacetime of more than four dimensions \cite{2, 3}. For some time, however, it was conventional to assume that in string  theory such extra dimensions were compactified to complex manifolds of small sizes   about the order of the  Planck length,  $\ell_P\sim ~ 10^{-33}$~cm \cite{3, 6},  or  they  could be even  of lower size independently of the Plank Length \cite{6a, 6b, 6d}; in this respect, the compactification process is a crucial step in the construction of models with extra dimensions \cite{5, 5t}.  \\
By taking into account the ideas explained above, in this paper we perform the Hamiltonian analysis of Proca's theory in 5D with a compact dimension. It is well know that four dimensional Proca's theory  is not a gauge theory, the theory describes a massive vector field and the physical degrees of freedom are three, this is, the addition of a mass term to Maxwell theory  breaks  the gauge invariance of the  theory and adds one physical degree of freedom to electromagnetic degrees of freedom \cite{5a, 5b}. Hence, in the present work, we study the effects of the compact extra dimension on a 5D Proca's theory.  Our study is based on a pure Dirac's analysis, this means that  we will develop all Dirac's steps in order to obtain a complete canonical analysis of the theory \cite{8a,8,9,9a}. We shall find the full constraints of the theory, the extended Hamiltonian and we will determine the full  Lagrange multipliers in order to construct the extended action.  Usually from consistency of the constraints it is not possible to determine the complete set of Lagrange multipliers, so a pure Dirac's analysis becomes relevant to determine all  them. Finally, we develop  the Hamiltonian study of a 5D  $BF$-like theory plus a massive term. We perform the campactactification process and we analyse the effective action. In particular, we show that the $BF$-like theory plus a massive  term  is a reducible system before and after  the compactification process.     All these ideas will be clarified along the paper. \\
The paper is organized as follows: In Sect. I,  we analyze a  Proca's theory in 5D, after performing  the compactification process on a  $S^1/\mathbf{Z_2}$ orbifold we obtain a  4D effective Lagrangian. We perform the Hamiltonian analysis and we obtain the complete constraints of the theory, the full Lagrange multipliers associated to the second class constraints and we construct the extended action. In addition, we carryout the counting of physical degrees of freedom. Additionally, in Sect. II, we perform the Hamiltonian analysis for a 5D  $BF$-like theory with a Proca's mass term; we also perform the compactification process on a $S^1/\mathbf{Z_2}$ orbifold,  and we  obtain  the effective action  after the compactification process. We show that for this theory there exist reducibility conditions among the first class constraints associated with the zero mode and the excited modes. Finally we carry out the counting of physical degrees of freedom. In Sect. III,  we  present some remarks and  prospects. 
 \newline
\newline
\section{Hamiltonian Dynamics for Proca theory in five dimensions  with a compact  dimension }
In this section, we shall perform the canonical analysis for Proca's  theory in five dimensions, then we will perform the compactfication process on  a $S^1/\mathbf{Z_2}$ orbifold. For this aim, the  notation that we will use along the paper  is the following: the capital latin indices $M, N$ run over $0,1,2,3, 5$,  here  as usual, $5$  label the compact dimension.  The $M, N$ indices  can be raised and lowered by the five-dimensional Minkowski metric $\eta_{M N}= (-1,1,1,1,1)$; $y$ will represent the coordinate in the compact dimension, $x^\mu$   the coordinates that label the points of  the four-dimensional manifold $M_4$  and $\mu, \nu=0,1,2, 3$ are spacetime indices; furthermore we will suppose  that  the compact dimension is a $S^1/\mathbf{Z_2}$ orbifold whose radius is $R$.\\
The Proca's Lagrangian in five dimensions without sources is given by 
\begin{eqnarray}\label{lp5}
L_{5p} = -\frac{1}{4}F_{MN}(x,y)F^{MN}(x,y)+\frac{m^{2}}{2}A_{M}(x,y)A^{M}(x,y),
\end{eqnarray}
where  $F_{MN}(x,y)=\partial_{M}A_{N}(x,y)-\partial_{N}A_{M}(x,y)$.\\
Because of the compactification of the fifth dimension will be carry out on a $S^1/\mathbf{Z_2}$ orbifold of  radius $R$,  such a choose imposes parity and  periodic conditions on the
gauge fields given by 
\begin{eqnarray}
A_{M}(x,y)&=&A_{M}(x,y+2\pi R), \nonumber \\
A_{\mu}(x,y)&=&A_{\mu}(x,-y),\nonumber \\
A_{5}(x,y)&=&-A_{5}(x,-y), 
\label{eq2}
\end{eqnarray}
thus,  the fields can be expanded in terms of  Fourier series as follows
\begin{eqnarray}\label{expancionA}
A_{\mu}(x,y)&=&\frac{1}{\sqrt{2\pi R}}A_{\mu}^{(0)}(x)+\sum_{n=1}^{\infty}\frac{1}{\sqrt{\pi R}}A_{\mu}^{(n)}(x)\cos{\left( \frac{ny}{R}\right) }, \nonumber \\
A_{5}(x,y)&=&\sum_{n=1}^{\infty}\frac{1}{\sqrt{\pi R}}A_{5}^{(n)}(x)\sin{\left( \frac{ny}{R}\right) }.
\end{eqnarray}
We shall  suppose that the number of KK-modes is $k$, and we will  take the limit $k \rightarrow \infty$  at the end of the calculations, thus,  $n=1, 2, 3...k-1$.
Moreover, by expanding the five dimensional Lagrangian  $L_{5p}$,  takes  the following form 
\begin{eqnarray}
L_{5p}(x,y)& = & -\frac{1}{4}F_{\mu\nu}(x,y)F^{\mu\nu}(x,y)+\frac{m^{2}}{2}A_{\mu}(x,y)A^{\mu}(x,y)-\frac{1}{2}F_{\mu 5}(x,y)F^{\mu 5}(x,y) \nonumber \\ 
& & +\frac{m^{2}}{2}A_{5}(x,y)A^{5}(x,y).
\label{eq4a}
\end{eqnarray}
Now, by inserting    (\ref{expancionA})  into (\ref{eq4a}),  and after performing  the integration on the  $y$ coordinate, we obtain the following 4D effective Lagrangian
\begin{eqnarray}\label{lp}
L_{p}(x) & = & \int \Big\{  -\frac{1}{4}F_{\mu\nu}^{(0)}(x)F^{\mu\nu}_{(0)}(x)+\frac{m^{2}}{2}A_{\mu}^{(0)}(x)A^{\mu}_{(0)}(x)\nonumber\\
&&+\sum_{n=1}^{\infty}\left[-\frac{1}{4}F_{\mu\nu}^{(n)}(x)F^{\mu\nu}_{(n)}(x)+\frac{m^{2}}{2}A_{\mu}^{(n)}(x)A^{\mu}_{(n)}(x)+\frac{m^{2}}{2}A_{5}^{(n)}(x)A^{5}_{(n)}(x)\right. \nonumber\\
&&\left. -\frac{1}{2}\left(\partial_{\mu}A_{5}^{(n)}(x)+\frac{m}{R}A_{\mu}^{(n)}(x) \right)\left(\partial^{\mu}A_{5}^{(n)}(x)+\frac{m}{R}A^{\mu(n)}(x) \right) \right]  \Big\} dx^4.
\label{eq4}
\end{eqnarray}
 The terms given by  $-\frac{1}{4}F_{\mu\nu}^{(0)}(x)F^{\mu\nu}_{(0)}(x)+\frac{m^{2}}{2}A_{\mu}^{(0)}(x)A^{\mu}_{(0)}(x)$ are called the  zero mode of the Proca theory \cite{5a, 5b}, and  the following terms are identified as a tower of KK-modes \cite{7}.  \\
In order to perform the Hamiltonian analysis, we observe that the theory is singular. In fact, it is straightforward to observe that the Hessian for the zero mode given by 
\begin{eqnarray*}
W^{\rho \lambda(0)}=\dfrac{\partial^{2}L_{p}}{\partial(\partial_{0}A_{\rho}^{(0)})\partial(\partial_{0}A_{\lambda}^{(0)})} & = & -\frac{g^{\mu\alpha}g^{\nu\beta}}{4}\left[\left(\delta_{\mu}^{0}\delta_{\nu}^{\lambda}-\delta_{\nu}^{0}\delta_{\mu}^{\lambda}\right)\dfrac{\partial F_{\alpha\beta}^{(0)}}{\partial(\partial_{0}A_{\rho}^{(0)})}+\left(\delta_{\alpha}^{0}\delta_{\beta}^{\lambda}-\delta_{\beta}^{0}\delta_{\alpha}^{\lambda} \right)\dfrac{\partial F_{\mu\nu}^{(0)}}{\partial(\partial_{0}A_{\rho}^{(0)})} \right]\\
& = & g^{\rho 0}g^{\lambda 0}-g^{\rho\lambda}g^{00}\\
&=& g^{\rho 0}g^{\lambda 0} + g^{\rho\lambda},
\end{eqnarray*}
 has $detW^{(0)}=0$,  rank$=3$ and one  null vector.  Furthermore, the Hessian of the $KK$-modes  has the following form 
 \begin{eqnarray*}
W^{HL(l)}&=&\dfrac{\partial^{2}L_{p}}{\partial(\partial_{0}A_{H}^{(l)})\partial(\partial_{0}A_{L}^{(l)})}\\
&=&g^{H0}g^{L0}-g^{HL}g^{00}-g^{00}\delta_{5}^{L}\delta_{5}^{H}\\
&=&g^{H0}g^{L0} + g^{HL} + \delta_{5}^{L}\delta_{5}^{H},
\end{eqnarray*}
and has $det W^{(l)}=0$, rank$=4k-4$ and $k-1$ null vectors.  Thus, a pure Dirac's method calls the definition of the  canonical momenta $( \pi_{(0)}, \pi_{(n)}^{i}, \pi_{(n)}^{5})$ to the dynamical variables  $(A_{\mu}^{(0)}, A_{\mu}^{(n)}, A_{5}^{(n)})$ given by 
\begin{eqnarray}\label{vA0}
\pi_{(0)}^{i}=-\partial^{i}A_{0}^{(0)}+\partial_{0}A^{i(0)},
\end{eqnarray}
\begin{eqnarray}\label{vAn}
\pi_{(n)}^{i}=-\partial^{i}A_{0}^{(n)}+\partial_{0}A^{i(n)},
\end{eqnarray}
\begin{eqnarray}\label{vA5}
\pi_{(n)}^{5}&=&\partial_{0}A_{5}^{(n)}+\frac{n}{R}A_{0}^{(n)}.
\end{eqnarray}
From the null vectors we identify the following  $k$ primary constraints 
\begin{eqnarray}\label{pi00}
\phi^{1}_{(0)}=\pi_{(0)}^{0}\approx 0,
\end{eqnarray}
\begin{eqnarray}\label{pi0n}
\phi^{1}_{(n)}=\pi_{(n)}^{0}\approx 0.
\end{eqnarray}
Hence,  the canonical   Hamiltonian is given by 
\begin{eqnarray}
H_{c}&=&\int \left[-A_{0}^{(0)}(x)\partial_{i}\pi^{i}_{(0)}(x)+\frac{1}{2}\pi^{i}_{(0)}(x)\pi_{i}^{(0)}(x)+\frac{1}{4}F_{ij}^{(0)}(x)F^{ij}_{(0)}(x)\right.\nonumber\\
&&-\frac{m^{2}}{2}A_{\mu}^{(0)}(x)A^{\mu}_{(0)}(x) + \sum_{n=1}^{\infty}\left(-A_{0}^{(n)}(x)\partial_{i}\pi^{i}_{(n)}(x)+\frac{1}{2}\pi^{i}_{(n)}(x)\pi_{i}^{(n)}(x)\right. \nonumber\\
&&+\frac{1}{2}\pi_{(n)}^{5}(x)\pi_{(n)}^{5}(x)-\frac{n}{R}\pi_{(n)}^{5}(x)A_{0}^{(n)}(x)+\frac{1}{4}F_{ij}^{(n)}(x)F^{ij}_{(n)}(x)\nonumber\\
&&-\frac{m^{2}}{2}A_{\mu}^{(n)}(x)A^{\mu}_{(n)}(x)-\frac{m^{2}}{2}A_{5}^{(n)}(x)A^{5}_{(n)}(x)\nonumber \\
&&\left.\left. +\frac{1}{2}\left(\partial_{i}A_{5}^{(n)}(x)+\frac{n}{R}A_{i}^{(n)}(x)\right)\left(\partial^{i}A_{5}^{(n)}(x)+\frac{n}{R}A^{i(n)}(x)\right) \right)\right]d^{3}x,
\end{eqnarray}
 by using the primary constraints, we identify the primary Hamiltonian
\begin{eqnarray}\label{h1p}
H_{p}&\equiv& H_{c}+\int\lambda_{1}^{(0)}(x)\phi_{(0)}^{1}(x)d^{3}x+\int\sum_{n=1}^{\infty}\lambda_{1}^{(n)}(x)\phi_{(n)}^{1}(x)d^{3}x,
\end{eqnarray}
where  $\lambda_{1}^{(0)}$ y $\lambda_{1}^{(n)}$ are Lagrange multipliers enforcing the constraints.\\
Hence, from the consistency of the constraints we find that 
\begin{eqnarray*}
\dot{\phi}_{(0)}^{1}(x)&=&\lbrace\phi_{(0)}^{1}, H_{1}\rbrace\\ 
&=&\partial_{i}\pi^{i}_{(0)}(x)+m^{2}A^{0}_{(0)}(x)\approx 0,
\end{eqnarray*}
and 
\begin{eqnarray*}\dot{\phi}_{(n)}^{1}(x)&=&\left\lbrace\phi_{(n)}^{1},H_{1}\right\rbrace \\
&=&\partial_{i}\pi_{(n)}^{i}(x)+m^{2}A^{0}_{(n)}(x)+\frac{n}{R}\pi_{(n)}^{5}(x)\approx 0.
\end{eqnarray*}
Therefore there are the following secondary constraints 
\begin{eqnarray}\label{sr0}
\phi_{(0)}^{2}(x)&=&\partial_{i}\pi^{i}_{(0)}(x)+m^{2}A_{(0)}^{0}(x)\approx 0,
\end{eqnarray}
\begin{eqnarray}\label{srn}
\phi_{(n)}^{2}(x)&=&\partial_{i}\pi^{i}_{(n)}(x)+\frac{n}{R}\pi^{5}_{(n)}(x)+m^{2}A_{(n)}^{0}(x)\approx 0.
\end{eqnarray}
On the other hand, the concistency  of  seconday constraints implies that 
\begin{eqnarray*}
\dot{\phi}_{(0)}^{2}(x)=m^{2}\partial_{i}A^{i}_{(0)}(x)-m^{2}\lambda_{1}^{(0)}(x)\approx 0,
\end{eqnarray*}
hence, 
\begin{eqnarray}\label{lambda0}
\lambda_{1}^{(0)}(x)\approx\partial_{i}A^{i}_{(0)}(x),
\end{eqnarray}
and
\begin{eqnarray*}\dot{\phi}_{(n)}^{2}(x)&=&m^{2}\partial_{i}A^{i}_{(n)}(x)-\frac{2n}{R}\partial_{i}\left(\partial^{i}A_{5}^{(n)}(x)+\frac{n}{R}A^{i(n)}(x)\right)-m^{2}\lambda_{1}^{(n)}(x)+\frac{m^{2}n}{R}A^{5}_{(n)}(x)\\
&\approx&0,
\end{eqnarray*}
thus
\begin{eqnarray}\label{lambdan}
\lambda_{1}^{(n)}(x)\approx\partial_{i}A^{i(n)}(x)-\frac{2n}{m^{2}R}\partial_{i}\left(\partial^{i}A_{5}^{(n)}(x)+\frac{n}{R}A^{i(n)}(x)\right)+\frac{n}{R}A^{5(n)}(x).
\end{eqnarray}
For this theory there are not third contraints.\\
By following with the method, we need to identify the first class and second class constraints. For this step we calculate the Poisson brackets among the primary and secondary constraints for the zero mode, obtaining 
\begin{eqnarray*}
\left(W^{'\alpha\beta(0)}\right)&=&\left(\begin{array}{cc}
\lbrace\phi_{(0)}^{1}(x),\phi_{(0)}^{1}(z) \rbrace & \lbrace\phi_{(0)}^{1}(x),\phi_{(0)}^{2}(z) \rbrace \\ 
\lbrace\phi_{(0)}^{2}(x),\phi_{(0)}^{1}(z) \rbrace & \lbrace\phi_{(0)}^{2}(x),\phi_{(0)}^{2}(z) \rbrace
\end{array} \right) \\
&=&\left(\begin{array}{cc}
0 & \lbrace\pi_{(0)}^{0}(x),\partial_{i}\pi^{i}_{(0)}(z)+m^{2}A^{0}_{(0)}(z)\rbrace \\ 
\lbrace\partial_{i}\pi^{i}_{(0)}(x)+m^{2}A^{0}_{(0)}(x),\pi_{(0)}^{0}(z) \rbrace & 0
\end{array} \right) \\
&=&\left(\begin{array}{cc}
0 & m^{2}\delta^{3}(x-z) \\ 
-m^{2}\delta^{3}(x-z) & 0
\end{array} \right)\\
&=&m^{2}\left(\begin{array}{cc}
0 & 1 \\ 
-1 & 0
\end{array} \right)\delta^{3}(x-z).
\end{eqnarray*}
The matrix  $ W^{\alpha \beta '(0)}$  has a rank= 2, therefore the constraints found for the zero mode are of second class.  In the same way, the Poisson brackets among the constraints  related for  the  $KK$-modes,  we find
\begin{eqnarray*}
\left(W^{'\alpha\beta(n)}\right)&=&\left(\begin{array}{cc}
\lbrace\phi_{(n)}^{1}(x),\phi_{(n)}^{1}(z) \rbrace & \lbrace\phi_{(n)}^{1}(x),\phi_{(n)}^{2}(z) \rbrace \\ 
\lbrace\phi_{(n)}^{2}(x),\phi_{(n)}^{1}(z) \rbrace & \lbrace\phi_{(n)}^{2}(x),\phi_{(n)}^{2}(z) \rbrace
\end{array} \right) \\
&=&\left(\begin{array}{cc}
0 & \lbrace\pi_{(n)}^{0}(x),m^{2}A^{0}_{(n)}(z)\rbrace \\ 
\lbrace m^{2}A^{0}_{(n)}(x),\pi_{(n)}^{0}(z) \rbrace & 0
\end{array} \right) \\
&=&\left(\begin{array}{cc}
0 & m^{2}\delta^{3}(x-z) \\ 
-m^{2}\delta^{3}(x-z) & 0
\end{array} \right)\\
&=&m^{2}\left(\begin{array}{cc}
0 & 1 \\ 
-1 & 0
\end{array} \right)\delta^{3}(x-z), 
\end{eqnarray*}
the matrix $W^{'\alpha\beta(n)}$ has a rank= $2(k-1)$, thus, the constraints associated to the $KK$-modes are of second class as well. In this manner, the counting of physical degrees of freedom is given in the following way: there are $10k-2$ dynamical variables, and $2k-2 + 2= 2k$ second class constraints, there are not first class constraints. Therefore, the number of physical degrees of freedom is $4k-1$. It is important to note that for $k=1$  we obtain the three  degrees of freedom  of a four dimensional Proca's  theory which  is   identified with the zero mode \cite{5a, 5b}.\\
Furthermore, we found  $2k$ second class constraints, which   implies that $2k$ Lagrange multipliers must be fixed;  however, we have found  only $k$ given in the expressions  (\ref{lambda0}) and   (\ref{lambdan}). Hence, let us  to find the full Lagrange multipliers;  it is important to comment  that  usually the Lagrange multipliers can be determined by means consistency conditions, however, for the theory under study this is not possible because  some of them  did not emerge  from the consistency of the constraints.  In order to construct the extended action and the extended Hamiltonian, we need identify all the  Lagrange multipliers, hence, for this important step, we can find the Lagrange multipliers by means of 
\begin{eqnarray}
\dot{\phi}^{\alpha}(x)=\left\lbrace\phi^{\alpha},H_{c}\right\rbrace+\lambda_{\beta} \left\lbrace\phi^{\alpha},\phi^{\beta}\right\rbrace\approx 0,
\label{eq16}
\end{eqnarray}
where $\phi^{\beta}$ are all the constrains found. In fact, by calling  $C^{\alpha\beta}=\left\lbrace\phi^{\alpha},\phi^{\beta}\right\rbrace$ and $h^{\alpha}=\left\lbrace\phi^{\alpha},H_{c}\right\rbrace$, we rewrite  (\ref{eq16}) as 
\begin{eqnarray}
h^{\alpha}+C^{\alpha\beta}\lambda_{\beta}\approx 0.
\end{eqnarray}
Therefore the Lagrange multipliers are given by  \cite{9a}
\begin{eqnarray}\label{lambda}
\lambda_\beta=-C^{-1}_{\beta \rho} h^{\rho}.
\end{eqnarray}
In this manner, for the zero mode we obtain
\begin{eqnarray*}
\left(C^{\alpha\beta(0)}\right)&=&\left( \begin{array}{cc}
\left\lbrace\phi_{(0)}^{1}(x),\phi_{(0)}^{1}(z)\right\rbrace & \left\lbrace\phi_{(0)}^{1}(x),\phi_{(0)}^{2}(z)\right\rbrace \\ 
\left\lbrace\phi_{(0)}^{2}(x),\phi_{(0)}^{1}(z)\right\rbrace & \left\lbrace\phi_{(0)}^{2}(x),\phi_{(0)}^{2}(z)\right\rbrace
\end{array} \right) \\
&=&m^{2}\left(\begin{array}{cc}
0 & 1 \\ 
-1 & 0
\end{array} \right)\delta^{3}(x-z),
\label{eq19a}
\end{eqnarray*}
and  its inverse is given by 
\begin{eqnarray*}
C^{(0)-1}_{\alpha \beta}=\frac{1}{m^{2}}\left(\begin{array}{cc}
0 & -1 \\ 
1 & 0
\end{array} \right)\delta^{3}(x-z),
\end{eqnarray*}
so,    for the constraints associated with the  zero modes,  $h^{(0)}$ is given by 
\begin{eqnarray*}
h^{(0)}&=&\left(\begin{array}{c}
\left\lbrace\phi_{(0)}^{1}(x),H_{c}\right\rbrace \\ 
\left\lbrace\phi_{(0)}^{2}(x),H_{c}\right\rbrace
\end{array} \right)\\
&=&\left(\begin{array}{c}
\partial_{i}\pi_{(0)}^{i}(x)+m^{2}A^{0}_{(0)}(x) \\ 
m^{2}\partial_{i}A^{i}_{(0)}(x)
\end{array} \right) ,
\end{eqnarray*}
therefore, by using  (\ref{lambda}) and (\ref{eq19a})  we can determine the Lagrange multipliers associated with the zero modes,  
\begin{eqnarray*}
\left( \begin{array}{c}
\lambda_{1}^{(0)}(x) \\ 
\lambda_{2}^{(0)}(x)
\end{array} \right) \approx -\frac{1}{m^{2}}\left(\begin{array}{cc}
0 & -1 \\ 
1 & 0
\end{array} \right)\left(\begin{array}{c}
\partial_{i}\pi_{(0)}^{i}(x)+m^{2}A^{0}_{(0)}(x) \\ 
m^{2}\partial_{i}A^{i}_{(0)}(x)
\end{array} \right)\delta^{3}(x-z),
\end{eqnarray*}
thus 
\begin{eqnarray}
\lambda_{1}^{(0)}(x)&\approx&\partial_{i}A^{i (0)}(x),\\\
\lambda_{2}^{(0)}(x)&\approx&-\frac{1}{m^{2}}\partial_{i}\pi^{i (0)}(x)-A^{0 (0)}(x).\label{lambda02}
\end{eqnarray}
In the same way, for the $KK$-modes we observe that 
\begin{eqnarray*}
\left(C^{\alpha\beta(n)}\right)&=&\left( \begin{array}{cc}
\left\lbrace\phi_{(n)}^{1}(x),\phi_{(n)}^{1}(z)\right\rbrace & \left\lbrace\phi_{(n)}^{1}(x),\phi_{(n)}^{2}(z)\right\rbrace \\ 
\left\lbrace\phi_{(n)}^{2}(x),\phi_{(n)}^{1}(z)\right\rbrace & \left\lbrace\phi_{(n)}^{2}(x),\phi_{(n)}^{2}(z)\right\rbrace
\end{array} \right)\\
&=&m^{2}\left(\begin{array}{cc}
0 & 1 \\ 
-1 & 0
\end{array} \right)\delta^{3}(x-z),
\end{eqnarray*}
where the inverse is
$$C^{(n)-1}_{\alpha \beta}=\frac{1}{m^{2}}\left(\begin{array}{cc} 
0 & -1 \\ 
1 & 0
\end{array} \right)\delta^{3}(x-z),$$
so, for the constraints associated with the excited modes,   $h^{(n)}$ is given by 
\begin{eqnarray*}
h^{(n)}&=&\left(\begin{array}{c}
\left\lbrace\phi_{(n)}^{1}(x),H_{c}\right\rbrace \\ 
\left\lbrace\phi_{(n)}^{2}(x),H_{c}\right\rbrace
\end{array} \right)\\
&=&\left(\begin{array}{c}
\partial_{i}\pi_{(n)}^{i}(x)+m^{2}A^{0}_{(0)}(x)+\frac{n}{R}\pi_{(n)}^{5}(x) \\ 
m^{2}\partial_{i}A^{i(n)}(x)-\frac{2n}{R}\partial_{i}\left(\partial^{i}A_{5}^{(n)}(x)+\frac{n}{R}A^{i(n)}(x) \right)+\frac{nm^{2}}{R}A^{5(n)}(x) 
\end{array} \right) ,
\end{eqnarray*}
additionally by using  (\ref{lambda}) we obtain 
\begin{eqnarray*}
\left( \begin{array}{c}
\lambda_{1}^{(n)}(x) \\ 
\lambda_{2}^{(n)}(x)
\end{array} \right)&=&-\left(\begin{array}{cc}
0 & -1 \\ 
1 & 0
\end{array} \right)\left(\begin{array}{c}
\partial_{i}\pi_{(n)}^{i}(x)+m^{2}A^{0(0)}(x)+\frac{n}{R}\pi_{(n)}^{5}(x) \\ 
m^{2}\partial_{i}A^{i(n)}(x)-\frac{2n}{R}\partial_{i}\left(\partial^{i}A_{5}^{(n)}(x)+\frac{n}{R}A^{i(n)}(x) \right)+\frac{nm^{2}}{R}A^{5(n)}(x) 
\end{array}\right)\\
&&\times\frac{\delta^{3}(x-z)}{m^{2}}, 
\end{eqnarray*}
thus, the Lagrange multipliers associated for  the second class constraints  of  the $KK$-modes read 
\begin{eqnarray}
\lambda_{1}^{(n)}(x)&=&\partial_{i}A^{i(n)}(x)-\frac{2n}{m^{2}R}\partial_{i}\left(\partial^{i}A^{5(n)}(x)+\frac{n}{R}A^{i(n)}(x)\right)+\frac{n}{R}A^{5(n)}(x) ,\\\
\lambda_{2}^{(n)}(x)&=&-\frac{1}{m^{2}}\partial_{i}\pi^{i}_{(n)}(x)+A_{0}^{(n)}(x)-\frac{n}{m^{2}R}\pi_{(n)}^{5}(x).
\end{eqnarray}
Hence, we have seen  that  by using a pure Dirac's method we were able to identify all Lagrange multipliers of the theory. In fact, Lagrange multipliers  are  essential in order to construct the extended  action. It is interesting  to point out  that in \cite{5a, 5b}  the complete Lagrange multipliers were not reported, thus, our approach extend the results reported in these works. \\
On the other hand, by using the matrix $C^{(0)-1}_{\alpha \beta}$ and $C^{(n)-1}_{\alpha \beta}$  it is straightforward to calculate the Dirac's brackets of the theory, thus,  we have a complete hamiltonian description of the system.  \\
By using all our results,  we are able to identify the extended action, hence by calling to the second class constraints as 
\begin{eqnarray*}
\chi_{(0)}^{1}(x)&\equiv&\pi_{(0)}^{0}(x)\approx0,\\\
\chi_{(0)}^{2}(x)&\equiv&\partial_{i}\pi_{(0)}^{i}(x)+m^{2}A_{(0)}^{0}(x)\approx0,\\\
\chi_{(n)}^{1}(x)&\equiv&\pi_{(n)}^{0}(x)\approx0,\\\
\chi_{(n)}^{2}(x)&\equiv&\partial_{i}\pi_{(n)}^{i}(x)+m^{2}A_{(0)}^{0}(x)+\frac{n}{R}\pi^{5}_{(n)}(x)\approx0, 
\end{eqnarray*}
now, by using the  second class constraints and the Lagrange multipliers  found for the zero mode and the excited modes, the extended action has the following expression 
\begin{eqnarray}\label{aeproca}
S_{E}[A,\pi,\bar{v}]&=&\int\left\lbrace \dot{A}_{\mu}^{(0)}\pi_{(0)}^{\mu}+A_{0}^{(0)}(x)\partial_{i}\pi^{i}_{(0)}(x)-\frac{1}{2}\pi^{i}_{(0)}(x)\pi_{i}^{(0)}(x)-\frac{1}{4}F_{ij}^{(0)}(x)F^{ij}_{(0)}(x)\right. \nonumber\\
&&+\frac{m^{2}}{2}A_{\mu}^{(0)}(x)A^{\mu}_{(0)}(x) - \bar{v}_{j}^{(0)}\chi_{(0)}^{j}+ \sum_{n=1}^{\infty}\left[\dot{A}_{\mu}^{(n)}\pi_{(n)}^{\mu} + \dot{A}_{5}^{(n)}\pi_{n}^{5}\right.\nonumber\\
&&\left. +A_{0}^{(n)}(x)\partial_{i}\pi^{i}_{(n)}(x)-\frac{1}{2}\pi^{i}_{(n)}(x)\pi_{i}^{(n)}(x)-\frac{1}{2}\pi_{(n)}^{5}(x)\pi_{(n)}^{5}(x)+\frac{n}{R}\pi_{(n)}^{5}(x)A_{0}^{(n)}(x)\right.\nonumber\\
&&\left.-\frac{1}{4}F_{ij}^{(n)}(x)F^{ij}_{(n)}(x)+\frac{m^{2}}{2}A_{\mu}^{(n)}(x)A^{\mu}_{(n)}(x)+\frac{m^{2}}{2}A_{5}^{(n)}(x)A^{5}_{(n)}(x)\right.\nonumber\\
&&\left. \left. -\frac{1}{2}\left(\partial_{i}A_{5}^{(n)}(x)+\frac{n}{R}A_{i}^{(n)}(x)\right)\left(\partial^{i}A_{5}^{(n)}(x)+\frac{n}{R}A^{i(n)}(x)\right)-\bar{v}_{j}^{(n)}\chi_{(n)}^{j}\right]\right\rbrace d^{4}x , 
\end{eqnarray}
where $\bar{v}_{j}^{(0)}$ and $\bar{v}_{j}^{(n)}$ are Lagrange multipliers enforcing the second class constraints. From the extended action, we are able to identify the extended Hamiltonian  given by
\begin{eqnarray}\label{heproca}
H_{E}&=&\int\left\lbrace A_{0}^{(0)}(x)\partial_{i}\pi^{i}_{(0)}(x)-\frac{1}{2}\pi^{i}_{(0)}(x)\pi_{i}^{(0)}(x)\right. \nonumber\\
&& -\frac{1}{4}F_{ij}^{(0)}(x)F^{ij}_{(0)}(x) +\frac{m^{2}}{2}A_{\mu}^{(0)}(x)A^{\mu}_{(0)}(x)\nonumber\\
&& + \sum_{n=1}^{\infty}\left[A_{0}^{(n)}(x)\partial_{i}\pi^{i}_{(n)}(x)\right.\nonumber\\
&& -\frac{1}{2}\pi^{i}_{(n)}(x)\pi_{i}^{(n)}(x)-\frac{1}{2}\pi_{(n)}^{5}(x)\pi_{(n)}^{5}(x) +\frac{n}{R}\pi_{(n)}^{5}(x)A_{0}^{(n)}(x)\nonumber\\
&& -\frac{1}{4}F_{ij}^{(n)}(x)F^{ij}_{(n)}(x)+\frac{m^{2}}{2}A_{\mu}^{(n)}(x)A^{\mu}_{(n)}(x)+\frac{m^{2}}{2}A_{5}^{(n)}(x)A^{5}_{(n)}(x)\nonumber\\
&&\left. \left. -\frac{1}{2}\left(\partial_{i}A_{5}^{(n)}(x)+\frac{n}{R}A_{i}^{(n)}(x)\right)\left(\partial^{i}A_{5}^{(n)}(x)+\frac{n}{R}A^{i(n)}(x)\right) \right]\right\rbrace d^{3}x .
\end{eqnarray}
It is worth  to comment that there are not first class constraints, therefore there is  not gauge symmetry; the system under study is not a gauge theory and we are able to observe from (\ref{eq4})  that the field $A_{\mu}^{(n)}$ is a massive vector field with a  mass term given by $ (m^2 + \frac{n^2}{R^2})$ and $A^{5}_{(n)}$ is a massive scalar field with a  mass term given by $m^2$.  \\
\section{Hamiltonian Dynamics for a Bf-like theory plus a Proca term   in five dimensions  with a compact  dimension }
In this section we shall analyze the following action 
\begin{equation}
S[A, B]= \int_M \left(  B^{MN} F_{MN} - \frac{m^2}{4} A_M A^M\right) dx^5,
\label{ac2}
\end{equation}
here $B^{MN}=-B^{NM}$ is an  antisymmetric field, and $A_M$ is the connexion. The Hamiltonian analysis of the $BF$-like term has been developed in \cite{10}, the theory is devoid  of physical degrees of freedom, the first class  constraints present reducibility conditions and the  extended Hamiltonian is a linear combination of first class constrains. Hence,  it is an interesting exercise to perform the analysis of the action (\ref{ac2}) in the context of extra dimensions. We expect that the massive term gives physical degrees of freedom to the full action and it  breaks down the general covariance of the theory.     \\
We shall  resume the complete Hamiltonian analysis of (\ref{ac2});   thus, we perform the $4+1$ decomposition, and  then we will carry out  the compactification  process on a $S^1/\mathbf{Z_2}$ orbifold   in order to obtain the following effective Lagrangian, 
\begin{equation}
L= B^{\mu \nu}_{(0)} F_{\mu \nu}^{(0)} - \frac{m^2}{4}A_{\mu} ^{(0)}A^{\mu}_{(0)} + \sum_{n=1}^{\mathcal{\infty}}\left[B^{\mu \nu}_{(n)} F_{\mu \nu}^{(n)} - \frac{m^2}{4}A_\mu^{(n)}A^{\mu}_{(n)}  + 2 B^{\mu 5} \left( \partial_\mu A_5^{(n)} + \frac{n}{R} A_\mu ^{n} \right)  \right].
\label{eq26}
\end{equation}
By performing the Hamiltonian analysis of the action (\ref{eq26}) we obtain the following results: there are  6 first class constraints for the zero mode 
\begin{eqnarray}
\gamma_{ij}^{(0)}&=& F_{ij}^{(0)} - \frac{1}{2} \left[\partial_i \Pi_{0j} ^{(0)}- \partial_j \Pi_{0i} ^{(0)}\right] \approx 0, \\ 
\gamma'{_{ij}}^{(0)}&=& \Pi_{ij}^{(0)} \approx0, 
\end{eqnarray}
here, $\left(\Pi^{(n)}_{MN}, \Pi^{M}_{(n)} \right)$ are canonically conjugate to $\left(B^{M N}_{(n)}, A^{(n)}_{M} \right)$ respectively. Furthermore,  these constraints are not independent because there exist  the reducibility condition,   $\partial_i \epsilon ^{ijk}\gamma_{jk}^{(0)}=0$;  thus, there are $[6-1]=5$ independent first class constraints for the zero mode. Moreover, there are  
8 second class constraints 
\begin{eqnarray}
\chi_{(0)}&=& \partial_i \Pi^i_{(0)} -\frac{m^2}{2} A_{0{(0)}} \approx0, \\
\chi^i _{(0)}&=& \Pi^i _{(0)}-2 B ^{0i}_{(0)} \approx0, \\
\chi^0_{(0)}&=& \Pi^0_{(0)}\approx0,\\
\chi^{0i}_{(0)} &=&  \Pi^{0i}_{(0)}\approx0, 
\end{eqnarray}
thus, with that information we carry out the  counting the physical degrees of freedom for the zero mode,  we find that there is  one physical degree of freedom. In fact, the massive term adds that degree of freedom to the theory, just like  Proca's term to Maxwell theory. \\
On the other hand, for the exited modes there are  $12k-12$ first class constraints given by 
\begin{eqnarray}
\gamma^{(n)}_i &=& \partial_i A_5^{(n)} + \frac{n}{R} A_i ^{(n)}- \left[ \partial_i \Pi_{05}^{(n)} + \frac{n}{2R} \Pi_{0i}^{(n)} \right] \approx0, \\
\gamma_{ij}^{(n)}&=& F_{ij}^{(n)} - \frac{1}{2} \left[\partial_i \Pi_{0j} ^{(n)}- \partial_j \Pi_{0i} ^{(n)}\right] \approx 0, \\ 
\gamma'{_{ij}}^{(n)}&=& \Pi_{ij}^{(n)} \approx0, \\
\gamma{_{i5}}^{(n)}&=& \Pi_{i5}^{(n)} \approx0, \\
\end{eqnarray}
however, also these constraints are not independent because there exist the following reducibility conditions; there are $k-1$ conditions given by   $\epsilon ^{ijk}\partial_i \gamma_{jk}^{(n)}=0$,  and $3(k-1)$ conditions  given by $\partial_{i}{\gamma}^{(n)}_{j}-\partial_{j}{\gamma}^{(n)}_{i}-\frac{n}{R}{\gamma}^{(n)}_{ij}=0$. Hence, there are $[(12k-12)- (4k-4)]=8k-8 $ independent first class constraints. Furthermore, there are  $10k- 10$ second class constraints 
\begin{eqnarray}
\chi^i _{(n)} &=& \pi^i_{(n)} - 2B^{0i}_{(n)} \approx 0, \\
\chi^0 _{(n)} &=& \pi^0_{(n)} \approx 0 \\
\chi^{(n)}_{0i} &=& \Pi_{0i}^{(n)}\approx 0, \\
\chi^{(n)}_{05} &=& \Pi_{05}^{(n)}\approx 0, \\ 
\chi_{(n)}^{5} &=& \Pi^{5}_{(n)}- B^{05}_{(0)}\approx 0, \\ 
\chi_{(n)} &=&\partial_i \pi^i_{(n)} +\frac{n}{R} \pi^5_{(n)} + \frac{m^2}{2} A^0_{(n)} \approx 0.
\end{eqnarray}
In this manner, by performing the counting of physical degrees of freedom we find that there are $2k-2$ physical degrees of freedom for the excited modes.  So, for the full theory  zero modes plus $KK$-modes,  there are $2k-1 $ physical degrees of freedom. Therefore, the theory present reducibility conditions among  the constraints of the zero mode and the constraints of the $KK$-modes.  We  observe from the first class constraints that the variable  $A_{\mu}^{(n)}$ has  a mass term given by $m^2$ and is not a gauge field.  On the other hand, $B_{i j} ^{(n)}$ is  a  massless  gauge field. 

\section{Conclussions and Prospects}
In this paper, we have developed the Hamiltonian analysis for a 5D Proca's theory in the context of extra dimensions. By performing the compactification process on an orbifold we obtained the complete canonical description of the theory. We obtained the complete set of constraints, the full Lagrange multipliers and the extended action. From our results we conclude that the theory is not a gauge theory, there are only second class constraints. Thus, 5D Proca's theory with a compact dimension,   describes the propagation of a massive  vector field associated with the zero mode plus  a tower of excited massive vector fields whose mass depend of the number of modes,  and a massive scalar field. It is remarkable  that after the compactification    process the theory is  not a gauge theory, the exited modes and the scalar field  are not gauge fields. Thus, the symmetry of the 5D Procas's theory is not affected by the compactification process. Moreover,  we carry out the counting of physical degrees of freedom,  in particular, our results reproduce those  ones known  for  Proca's theory  without a compact dimension. Finally, in order to construct the extended action, we had  identified  the complete set of Lagrange multipliers; it is important to remark that usually Lagrange multipliers emerge from consistency condition of the constraints. However, if the Lagrange multipliers that emerge from consistency conditions  are mixed, then it is  difficult to determine them from constancy conditions. In those  cases, it is necessary to use the method performed in this paper;  thus, all Lagrange multipliers can be determined. \\
 On the other hand, we develop the Hamiltonian analysis of a 5D $BF$-like theory plus a massive Proca's term. From our analysis, we conclude that the theory  is not topological anymore. In fact, the massive term breakdown   the topological structure of the $BF$-like term. The theory present first and second class constraints, in order to carry out the correct counting of degrees of freedom, we identified  reducibility conditions among the first class constraints associated with the zero and the exited modes. The effect of the  massive term  is that add degrees of freedom to the topological $BF$-like term, just like Proca's term add degrees of freedom to Maxwell theory. Hence,  in order to study the quantization of the theories under study, we have in this paper all the necessary ingredients. It is important to comment that our results  can be extended to models that generalize  the dynamics of Yang-Mills theory,  as for instance,  the models  reported  in \cite{11}. In fact, in \cite{11} there are models involving topological theories and massive terms that generalize the Yang-Mills dynamics in three and four dimensions, in this respect, our work can be useful to study those models in the context of extra dimensions.

\noindent \textbf{Acknowledgements}\\[1ex]
This work was supported by Sistema Nacional de Investigadores M\'exico. The authors  want   to thank  R. Cartas-Fuentevilla for reading the manuscript.

\end{document}